\documentclass[aps,prb,preprint, superscriptaddress,showpacs]{revtex4}
\usepackage{graphicx}
\usepackage{dcolumn} 
\usepackage{bm}      
\newcommand{\bhf}{{$B_{hf}~$}}
\newcommand{\fe}{{$FeCl_4$~}}

\newcommand{\lb}{{$\lambda-(BETS)_2FeCl_4$~}}

\begin{document}

\preprint{APS/123-QED}

\title{M\"{o}ssbauer spectroscopy study of the ``mysterious" magnetic transition in $\lambda-(BETS)_2FeCl_4$}

\author{J. C. Waerenborgh}
\affiliation{Instituto Tecnol\'{o}gico  e Nuclear / CFMCUL,
Estrada Nacional  10, P-2686-953 Sacav\'{e}m, Portugal}
\author{S. Raba\c{c}a}
\affiliation{Instituto Tecnol\'{o}gico  e Nuclear / CFMCUL,
Estrada Nacional  10, P-2686-953 Sacav\'{e}m, Portugal}

\author{M. Almeida}
\affiliation{Instituto Tecnol\'{o}gico  e Nuclear / CFMCUL,
Estrada Nacional  10, P-2686-953 Sacav\'{e}m, Portugal}

\author{A. Kobayashi}
\affiliation{Department of Chemistry, College of Humanities and
Sciences, Nihon University, Sakurajosui 3-25-40, Setagaya-Ku,
Tokyo 156-8550, Japan}

\author{B. Zhou}
\affiliation{Department of Chemistry, College of Humanities and
Sciences, Nihon University, Sakurajosui 3-25-40, Setagaya-Ku,
Tokyo 156-8550, Japan}

\author{J.S. Brooks}
\affiliation{Department of Physics/Chemistry and National High
Magnetic Field Laboratory, Florida State University, Tallahassee,
FL 32310, USA}

\date{\today}

\begin{abstract}

The compound  $\lambda-(BETS)_2FeCl_4$ provides an effective
demonstration of the interaction of $\pi$ conduction electron and
d-electron localized moment systems in molecular crystalline
materials where antiferromagnetic insulating and magnetic field
induced superconducting states can be realized. The
metal-insulator transition has been thought to be cooperative,
involving both the itinerant $\pi$- electron and localized
d-electron spins where antiferromagnetic order appears in both
systems simultaneously. However, recent specific heat data has
indicated otherwise [Akiba \emph{et al.}, J. Phys. Soc. Japan
{\bf78},033601(2009)]: although the $\pi$-electron system orders
antiferromagnetically and produces a metal-insulator transition, a
``mysterious" paramagnetic d-electron state remains. We
report $^{57}Fe$ M\"{o}ssbauer measurements that support the paramagnetic model,
provided the d-electron spins remain in a fast relaxation state below the transition. From the measured hyperfine fields, we also determine the temperature dependence of the $\pi-d$ electron exchange field.
\end{abstract}

\maketitle

\lb (BETS = bisethylenedithio-tetraselenafulvalene) is one of the
most thoroughly studied molecular conductors in the last few years
due to its unique properties derived from the interaction between
conducting $\pi$-electrons in the BETS donor layers and localized
d-electrons in $FeCl_4$ anions with S=5/2
spins.\cite{KobayashiChemRev} The crystal structure of \lb as
shown in Fig. 1  consists in stacks of BETS donors along $a$ and tightly packed in layers with a 2D network of S...S contacts  parallel to $(a,c)$ , alternating along $b$ with layers of $FeCl_4$ anions. At high temperatures this compound is a quasi 2D
metal due to delocalization of $\pi$-electrons in the layered
network of partially oxidized donors, and its magnetic
susceptibility is dominated by  the paramagnetic S=5/2 $FeCl_4$
spins. At 8.3 K this compound undergoes a transition towards an antiferromagnetic (AF)
insulating ground state (see inset of Fig. 3 below). Since the
isomorphous compound with diamagnetic $GaCl_4$ anions remains
metallic, becoming superconductor at 6 K, the metal-insulator
transition in \lb has been thought to be driven by the ordering of the
Fe spins.

However until now no direct microscopic measurements
directly probing the role of the anions have been published and
recent specific heat measurements by Akiba et al. have cast some
doubts on the role of the S=5/2 \fe spins in the
transition.\cite{Akiba} These authors have suggested that during
the transition, while the $\pi$ spins order antiferromagnetically, the Fe spins remain
paramagnetic below 8.3K. According to their model an effective
field $H_{\pi-d}~ \approx$ 4 T caused by the ordering of the $\pi$
spin system is switched on at the Fe sites at approximately 8.3 K
but the $Fe^{3+}$  cations remain paramagnetic with the 3d energy
levels described by a Zeeman splitting. The latter gives rise to a
Schottky 6-level term in the specific heat.

\begin{figure}[tbp]
\linespread{1}
\par
\includegraphics[scale=.35,angle=0]{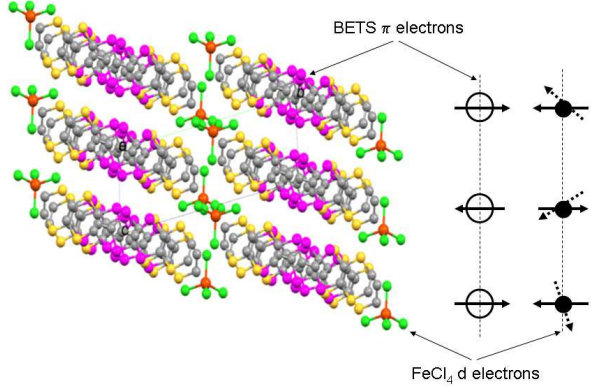}
\par
\caption{Crystal structure of \lb projected in the b-c plane(after
Ref. \cite{KobayashiJACS}). The charge transfer of one
electron between each two BETS donors and one $FeCl_4$ anion leaves a
delocalized spin 1/2 $\pi$  electron on the donor stacks, and a
localized d-electron at the anion site. The schematic shows the two possible spin
configurations below $T_N$ with fully antiferromagnetic $\pi$ and d order (solid arrows)
or with a paramagnetic d state (dashed arrows).} \label{Fig1}
\end{figure}

In this Communication we describe the results of a $^{57}Fe$
M\"{o}ssbauer spectroscopy study to examine the role
of the Fe S=5/2 spins in the transition. The single crystals of \lb used in this work were  grown using standard electrochemical methods  from 99\% $^{57}Fe$ enriched $TEAFeCl_4$.  M\"{o}ssbauer
spectra were collected with the absorber within a liquid-He bath
cryostat, in transmission mode using a conventional
constant-acceleration spectrometer and a 25 mCi $^{57}Co$ source
in a Rh matrix. The absorber was prepared by randomly placing
between two perspex plates approximately 4 mg of \lb single
crystals  99\% enriched in $^{57}Fe$.  Electrical transport measurements, which verified the
metal-insulator transition in the $^{57}$Fe enriched samples at
$T_N$ = 8.3 K (inset of Fig. 3), were carried out using a standard
4-terminal resistance configuration on crystals that had been used
in the M\"{o}ssbauer measurements.

\begin{figure}[tbp]
\par
\includegraphics[scale=.25,angle=0]{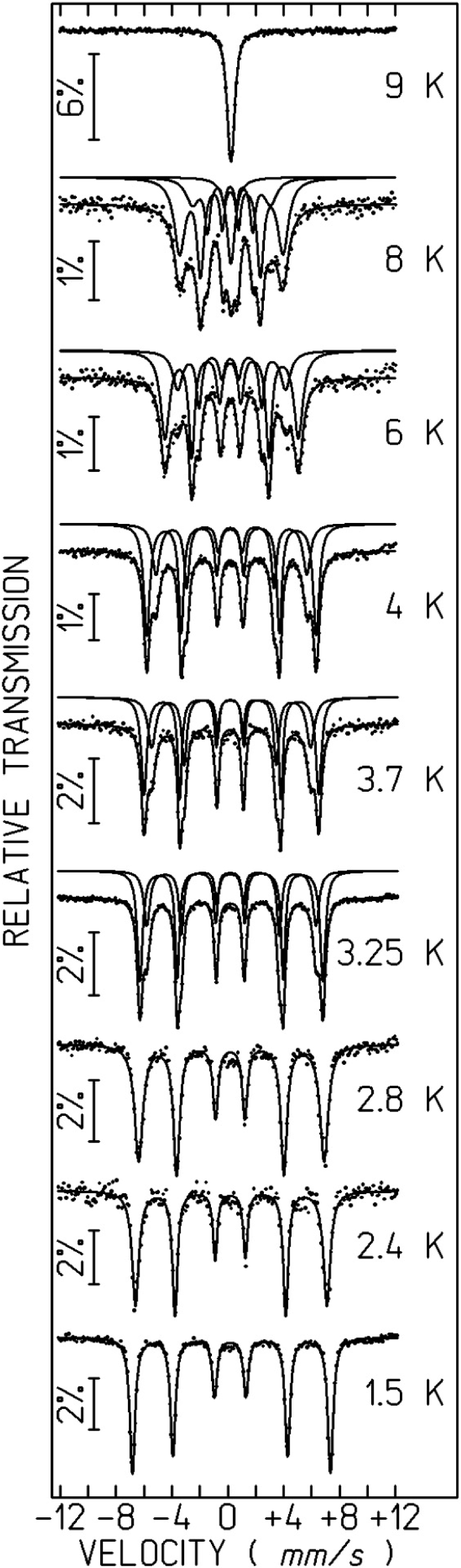}
\par
\caption{ $^{57}Fe$ M\"{o}ssbauer spectra of \lb for different
temperatures. The individual solid lines are components of the
spectrum coming from the two sextets and quadrupole components.
The solid line through the data (dots) is the sum of the
individual contributions. } \label{Fig2}
\end{figure}

\begin{figure}[tbp]
\linespread{1}
\par
\includegraphics[scale=.4,angle=0]{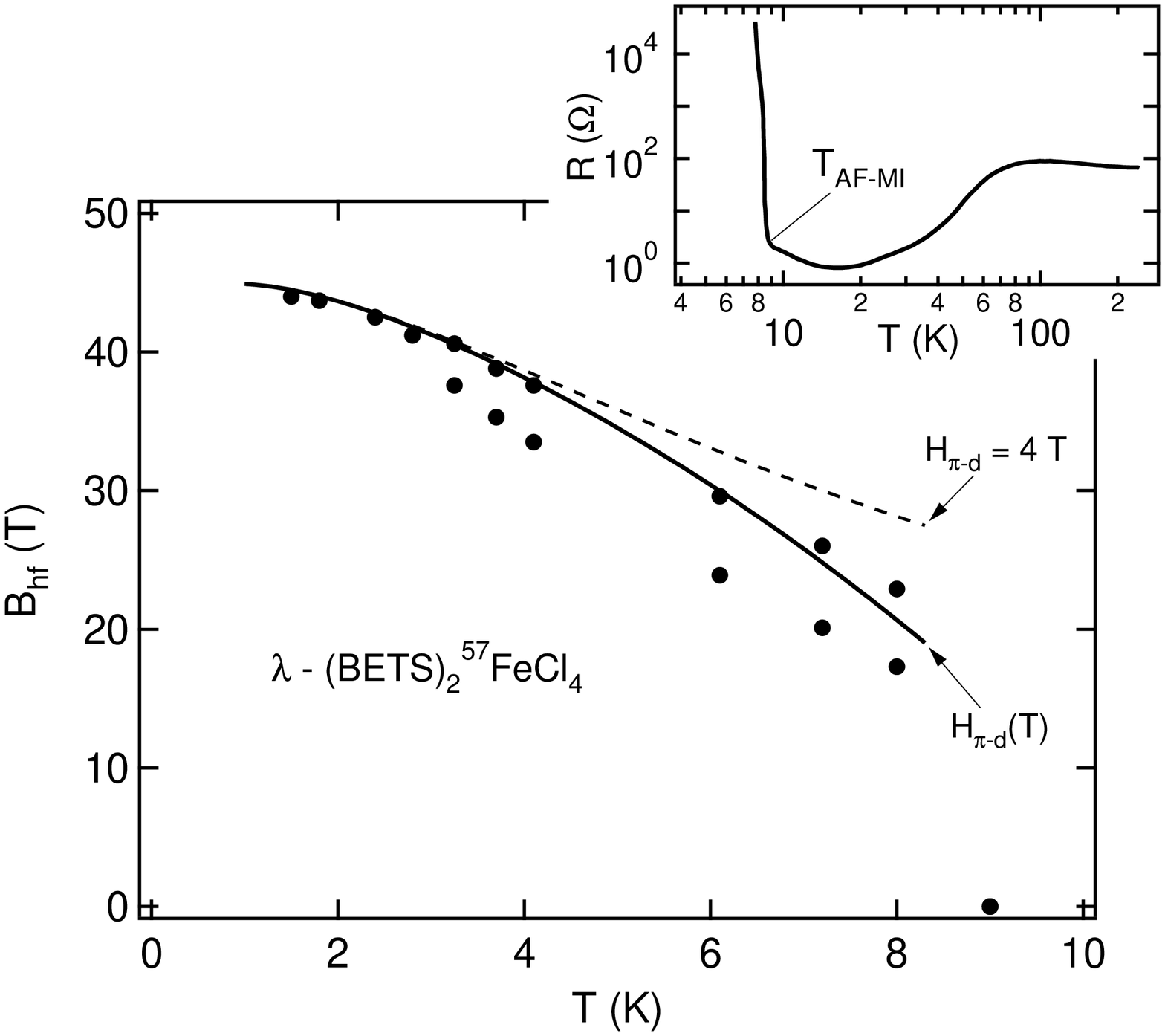}
\par
\caption{Temperature dependence of the Fe hyperfine fields,
$B_{hf}$ in \lb. The lines represent the fast relaxation model for both fixed and temperature dependent $H_{\pi-d}$ exchange fields (see text). Inset:
temperature dependent resistance of a single crystal of
$^{57}Fe$ enriched \lb showing the M-I transition at 8.3 K.}
\label{Fig3}
\end{figure}

 M\"{o}ssbauer spectra obtained at different temperatures
in the range 9 to 1.5 K are shown in Fig. 2, and the fitting
parameters are listed in Table 1. The spectra were fitted to Lorentzian lines using a
non-linear least-squares method. \cite{Rodrigues} The relative
areas and line widths of both peaks in a quadrupole doublet and
the peak pairs 1-6, 2-5 and 3-4 in a magnetic sextet were
constrained to remain equal during the refinement procedure.
Isomer shifts (IS) are given relative to $\alpha-Fe$ at room
temperature. Above 8.3 K the spectra present
a single line typical of paramagnetic $Fe^{3+}$ in a site with a
low quadrupole distortion (high symmetry environment). Below 8.3 K
sextet splittings are observed. The spectrum obtained at 8 K still
shows a small fraction (ca. 13\%) of ``paramagnetic" atoms, which
can be ascribed to a hysteresis or the slow dynamics of the
transition. Below this temperature, down to 3.2 K, two magnetic
splittings are clearly seen. They have identical isomer shifts,
but slightly different hyperfine fields, \bhf. This clearly
indicates that there are two different magnetic environments for
the Fe sites. The uncertainty in the relative areas of the sextets
is relatively large, but in a first approximation they present a
temperature independent 1:2 ratio. Below 3 K, these two sextets
appear to merge and only one sextet is observed in the range
2.8-1.5 K, suggesting a transition occurring at $3.0~ \pm~ 0.1 K$.
The hyperfine fields \bhf, shown in Fig. 3, present significant
temperature dependence until circa 3 K. Below this temperature they
seem to saturate approaching 44 T at 1.5 K, very close to the
value of \bhf=45 T observed for $FeCl_3$ at 4.2 K.\cite{Greenwood}

 The observed change of the shape of the M\"{o}ssbauer
spectra in a narrow temperature range around 8.3 K is indicative of
a magnetic ordering process.  Clearly, above the M-I transition the $Fe^{3+}$ spins are
oscillating with a relaxation frequency $\omega_R > 10^{12}~Hz$,
typical of paramagnetic $Fe^{3+}$. In this situation the magnetic
field observed at the Fe nuclei is averaged to
zero.\cite{Greenwood, Wickman68} In a first approximation the presence of sextets with sharp peaks as those observed below the M-I transition suggests that the Fe nuclei are
feeling a static magnetic hyperfine field within the observation
time scale of the M\"{o}ssbauer effect. This would imply that the
relaxation frequency of the $Fe^{3+}$  spins is now $\omega_R <
10^{8}~Hz$.

In the proposed picture of Akiba et al. \cite{Akiba} based on
specific heat data of \lb, the Fe S=5/2 spins remain paramagnetic
below the transition at 8.3 K, although subject to an internal
effective field $H_{\pi-d}$ $\sim$ 4 T switched on at the Fe sites by
the ordering of the $\pi$ spin system. In the \underbar{absence} of $H_{\pi-d}$, if the Fe atoms remain in
the paramagnetic state in a fast relaxation regime, the magnetic
field in the nuclei averages to zero and only a two-line pattern
should appear corresponding to the electric quadrupole interaction
(in this case a single absorption peak is observed due to the very
low quadrupole splitting), evident at 9 K. If there was a drastic
slowing down of the electronic relaxation, but in the absence of
magnetic ordering, two different sextets should be observed,
corresponding to the different $M_S$ states 5/2, and 3/2, as it has
been observed in diluted $^{57}$Fe doped compounds, such as in
Fe-doped $Al_2O_3$, \cite{Wickman66}$LiAl_5O_8$ \cite{Viccaro} or
in proteins such as the transferrins.\cite{Kretchmar} The $M_S$=1/2
state should result in a complicated 11-line pattern since it
induces non-diagonal terms in the Hamiltonian of the hyperfine
interactions.\cite{Pfannes} Although its presence was reported in
the transferrin case\cite{Kretchmar} it usually is not observed
due to enhanced relaxation of the $M_S$ = $\pm$ 1/2 electronic
doublet. \cite{Wickman66}

Unlike dilute systems, the concentration of $Fe^{3+}$  in \lb
is not so low as in the above examples and therefore due to
spin-spin interactions a slow relaxation regime is not expected to
occur. It could be argued that in \lb  the internal effective
field of the donors at the Fe sites, estimated as $H_{\pi-d}$  $\sim$ 4
T, could freeze the spin flipping and bring the system to a slow
relaxation regime. However there is no evidence for the $M_S$ = 3/2
state. The sextet with the smaller magnetic splitting observed
between 8 and 3K cannot correspond to the $M_S$ = 3/2 state because
its \bhf value is $\geq$ 90\% of the \bhf of the larger magnetic
splitting (below 4.1 K where thermal excitations are less
important). The saturated \bhf values associated with $M_S$ =3/2 and
5/2 electronic states should be proportional to $M_S$.
\cite{Kretchmar} Furthermore in such case the relative intensities
of both magnetic splittings should follow the evolution of their
statistical thermal population, \cite{Pfannes} while they remain
approximately constant.

The origin of the transition observed at approximately 3 K, as the
merging of the two sextets with only one hyperfine field is not
entirely clear, but may be associated with a change of the
magnetic wave vector in the AF state. Although not as dramatic as
in the M\"{o}ssbauer data, changes below $T_N$ have been seen in
other independent studies  in the 3 K region. Matsui and
co-workers\cite{MatsuiBETS} have investigated the microwave cavity
response with \lb, and below $T_N$ have found highly dispersive
modes attributed to charge degrees of freedom. However, for
$H_{ac}\parallel H\parallel a^* $, a peak in the cavity
dissipation ($\Delta\Gamma /2f_0$) appears at 3 K (H = 0) which,
due to the unfavorable direction of the eddy currents for
$H_{ac}\parallel$ a*, the authors attribute to a dynamic response
due to spin degrees of freedom. This peak has a complicated
dependence on H near the 1.2 T spin-flop transition. Likewise,
Rutel et al.\cite{Rutel} have observed anomalies in the microwave
cavity response below 4 K for $H\parallel c$.\cite{Rutel}

\newpage
Table I. Computed parameters from the M\"{o}ssbauer
spectra of \lb taken at different temperatures.

\begin{tabular}{|c|c|c|c|c|c|c|}
  \hline
  T  & IS  & QS, $\epsilon$  & $B_{hf}$  & $\Gamma$ & I \\
  (K) & (mm/s) & (mm/s) & (T) & (mm/s) &\\
  \hline
  200 & 0.27 & 0.21 & - & 0.49 & 100\%\\ \hline
  9.0 & 0.34 & 0.20 & - & 0.54 & 100\%\\ \hline
  8.0 & 0.34 & 0.20 & - & 0.57 & 13$\pm$2\%\\
      & 0.33 & 0.10 & 17.3 & 0.33 & 26$\pm$4\%\\
      & 0.33 & 0.07 & 22.9 & 0.40 & 61$\pm$4\%\\ \hline
  7.2 & 0.34 & 0.11 & 20.1 & 0.41 & 32$\pm$3\%\\
      & 0.34 & 0.08 & 26.0 & 0.39 & 68$\pm$3\%\\ \hline
  6.1 & 0.34 & 0.10 & 23.9 & 0.47 & 32$\pm$2\%\\
      & 0.34 & 0.10 & 29.6 & 0.52 & 68$\pm$2\%\\ \hline
  4.1 & 0.34 & 0.08 & 33.5 & 0.43 & 36$\pm$1\%\\
      & 0.34 & 0.09 & 37.6 & 0.42 & 64$\pm$1\%\\ \hline
  3.7 & 0.34 & 0.07 & 35.3 & 0.29 & 43$\pm$3\%\\
      & 0.34 & 0.08 & 38.8 & 0.33 & 57$\pm$3\%\\ \hline
  3.25 & 0.33 & 0.08 & 37.6 & 0.32 & 38$\pm$1\%\\
      & 0.33 & 0.08 & 40.6 & 0.36 & 62$\pm$1\%\\ \hline
  2.8 & 0.33 & 0.09 & 41.2 & 0.45 & 100\%\\ \hline
  2.4 & 0.33 & 0.07 & 42.5 & 0.38 & 100\%\\ \hline
  1.8 & 0.34 & 0.09 & 43.7 & 0.61 & 100\%\\ \hline
  1.5 & 0.34 & 0.09 & 44.0 & 0.44 & 100\%\\ \hline
\end{tabular}

\noindent IS, \textit{isomer shift relative to metallic Fe at 298 K.}

\noindent QS, \textit{quadrupole splitting.}

\noindent \textit{$\epsilon = (e^2V_{ZZ}Q/4)(3cos^2\theta - 1)$,
quadrupole shift.}

\noindent $B_{hf}$, \textit{magnetic hyperfine field.}

\noindent $\Gamma$, \textit{half-width of the doublet peaks.}

\noindent I, \textit{relative area.}

\noindent Estimated errors:\textit{ $\leq$ 0.002 mm/s for IS, QS,
$\epsilon$, $\Gamma$;} \noindent \textit{$\leq$ 0.2 T for
$B_{hf}$}

\bigskip
It is however difficult for a M\"{o}ssbauer probe to discriminate between the onset of magnetic order, spin-glass, behavior, or a particular case of a paramagnetic ``fast relaxation" behavior where the Fe spins are Zeeman split by an applied field and the population of M$_S$ states are different. The present results may therefore be  consistent with the Fe atoms remaining paramagnetic in the low temperature  state below
8.3 K. Assuming a fast relaxation model below the transition, we may compute the hyperfine field based on the Fe cations, inducing a Zeeman splitting and a Boltzmann distribution of the 6 M$_s$ states:

\begin{equation}
B_{hf}(T) = \sum_{M_s}[B(M_s)exp(X_{M_s})]/\sum_{M_s}exp(X_{M_s})
\label{eq:1}
\end{equation}

\noindent Here $X_{M_s}= -g\mu_BM_sH_{\pi-d}/k_BT$. We take $B(M_s)~\sim~\pm$45 T for M$_s$=$\mp$5/2, $\pm$27 T for M$_s$=$\mp$3/2, and $\pm$9 T for M$_s$=$\mp$1/2 (in proportion to M$_s$). In Fig. 3 the computed temperature dependence of $B_{hf}$ is shown for both a constant $H_{\pi-d}$ = 4 T, and for a temperature dependent $H_{\pi-d}$, increasing from 2.45 T at 8 K to 4.2 T at 1.5 K. We find that the best fit to the data implies that $H_{\pi-d}$ is temperature dependent. Although this modifies the temperature dependence of the Schottky specfic heat described by Akiba et al.\cite{Akiba}, the differences are not significant.

The temperature dependence of the exchange field may be described by a spin-wave behavior\cite{KittelEd3,SpinWave} where, for anti-ferromagnetic dispersion in the spin 1/2 $\pi$-electron system $\omega = J/\hbar|ka|$, $H_{\pi-d}(T)$= $H_{\pi-d}(0)(1-AT^{3})$. We find that for $A ~\sim~7 \times 10^{-4} K^{-3}$, Eq. 1 provides a reasonable description of the temperature dependence of $B_{hf}$, shown as the solid line in Fig. 3. From $A$ we estimate the spin-wave exchange energy to be $J~\sim~5.6 K$, comparable to $T_N$.

The suggestion and evidence that the magnetic order appears in the $\pi$-electron system, but not in the d-electron system, seems unusual\cite{UjiComment}. However, estimates do show that the mean-field exchange interaction of the $\pi$-electron system is the largest: $J_{\pi-\pi}$, $J_{\pi-d}$, and $J_{d-d}$ are 448, 14.6, and 0.64 K respectively\cite{Mori02}. Nevertheless, the d-electron spins must play a central role in the formation of the magnetic ground state. A  temporal probe of the spin dynamics of the d-electron system below the M-I transition, as well as  magnetic field dependent specific heat and
M\"{o}ssbauer experiments, would be useful to further explore the
nature of the magnetic order associated with this very
unusual antiferromagnetic metal-insulator transition.

\begin{acknowledgments}
This work was supported  by FCT(Portugal) through project PTDC/QUI/64967/2006 , and in part by NSF DMR-0602859 (JSB). The
National High Magnetic Field Laboratory is supported by NSF
DMR-0654118, by the State of Florida, and the DOE. The collaboration between Portuguese and American team members was supported by FLAD. The authors are indebted to Pedro Schlottmann and Kun Yang for valuable discussions.
\end{acknowledgments}




\begin{thebibliography}{9}
\bibitem{SpinWave} Here we assume that  $H_{pi-d}$ at the Fe site follows the fractional change of staggard magnetization in the $\pi$-electron system. Following Ref. \cite{KittelEd3} for linear antiferromagnetic dispersion, the $T^3$ prefactor is  $A = 2.40411(2\pi^2/SQ)(k_B/2JS)^3$ where S, J, and Q are the spin, exchange energy, and Fe sites per unit cell respectively, and the pre-factor is the value of the definite integral.
\end{thebibliography}
\end{document}